\documentclass[12pt,journal,onecolumn]{IEEEtran}
\usepackage{amsmath}
\usepackage{amssymb}
\usepackage{epsfig}
\usepackage{url}
\usepackage{color}
\usepackage{cite}
\usepackage{epsf}

\newtheorem{thm}{\textbf{Theorem}}[section]

\newtheorem{lem}[thm]{\textbf{Lemma}}

\renewcommand{\baselinestretch}{1.5}

\makeatletter  %% this is crucial
 \renewcommand\subsubsection{\@startsection{subsubsection}{3}{\z@}%
                        {-18\p@ \@plus -4\p@ \@minus -4\p@}%
                        {8\p@ \@plus 4\p@ \@minus 4\p@}%     <-- this is copied from the subsection command
                        {\normalfont\normalsize\bfseries\boldmath
                         \rightskip=\z@ \@plus 8em
 \pretolerance=10000 }}
\makeatother   %% this is crucial

\begin{document}

\title{Multilevel Coding over Two-Hop Single-User
Networks}
\author{ Vahid Pourahmadi, Alireza Bayesteh,
and~Amir~K.~Khandani\\
\small Coding \& Signal Transmission Laboratory (www.cst.uwaterloo.ca)\\
Dept. of Elec. and Comp. Eng., University of Waterloo\\ Waterloo, ON, Canada, N2L 3G1 \\
Tel: 519-725-7338, Fax: 519-888-4338\\e-mail: \{vpourahm, alireza,
khandani\}@cst.uwaterloo.ca\\
\thanks{Financial supports provided by Nortel, and the corresponding matching
 funds by the Federal government: Natural Sciences and Engineering Research
 Council of Canada (NSERC) and Province of Ontario: Ontario Centres of
 Excellence (OCE) are gratefully acknowledged.}}

\maketitle

\begin{abstract}  \label{sec:Abs}
In this paper, a two-hop network in which information is transmitted
from a source via a relay to a destination is considered. It is
assumed that the channels are static fading with additive white
Gaussian noise. All nodes are equipped with a single antenna and the
Channel State Information (CSI) of each hop is not available at the
corresponding transmitter. The relay is assumed to be simple, i.e.,
not capable of data buffering over multiple coding blocks,
water-filling over time, or rescheduling. A commonly used design
criterion in such configurations is the maximization of the {\em
average received rate} at the destination. We show that using a
continuum of multilevel codes at both the source and the relay, in
conjunction with decode and forward strategy at the relay, performs
optimum in this setup. In addition, we present a scheme to optimally
allocate the available source and relay powers to different levels
of their corresponding codes. The performance of this scheme is
evaluated assuming Rayleigh fading and compared with the previously
known strategies.
\end{abstract}

\section{Introduction} \label{sec:Intro}

In recent years, relay-assisted transmission has gained significant
attention as a powerful technique to enhance the performance of
wireless networks. The main idea is to employ some extra nodes
(relay nodes) in the network to facilitate the communication between
the terminal nodes. The concept of relaying was first introduced by
Van der Meulen in \cite{vandermeulen} and is defined as a scheme to
improve the coverage/reliability of a wireless network. For
instance, relays are usually deployed in networks when the direct
link between the source and the destination is either blocked or has
a very poor quality. The term \emph{two-hop} network usually refers
to such a network configuration in which there is no direct link
between the source and the destination, and one relay node assists
the transmission of data between the end terminals, see
Fig.~\ref{fig:NetModel}. Two-hop networks have been implemented
widely in different applications, including TV broadcasting and
satellite communications.

Following the introduction of relay channel in \cite{vandermeulen},
Cover and El Gamal introduce two different coding strategies for
single relay networks \cite{CoverRelay}. In the first strategy,
known as ``Decode and Forward'' ($\mathbf{DF}$), the relay decodes
the transmitted message and cooperates with the source to send the
message in the next block. Instead of decoding, in the second
strategy, the relay compresses the received signal and forwards it
to the destination in the next block. The terms ``Compress and
Froward'' ($\mathbf{CF}$) or ``Quantize and Forward''
($\mathbf{QF}$) usually refer to this transmission scheme. Besides
$\mathbf{DF}$ and $\mathbf{CF}$, in some recent results,
\cite{gallager,gastpar,zahedi1,zahedi2}, the authors investigate
another transmission scheme called ``Amplify and Forward''
($\mathbf{AF}$) for the Gaussian relay network. In this strategy,
without decoding the information, the relay amplifies the received
signal and retransmits it to the destination.

Knowing these schemes, the performance of relaying is analyzed for
different network topologies. For instance, considering a
single-relay network, authors of \cite{zahedi1} and \cite{zahedi2}
derive a single-letter expression for the  maximum achievable rate
of  $\mathbf{AF}$ relaying using a simple linear scheme (assuming
frequency division and AWGN channel). As another example,
\cite{gastpar}  shows that $\mathbf{AF}$ relaying achieves the
network capacity of Gaussian parallel single-antenna relay network.
The extension of \cite{gastpar} to the case of multiple-antenna
Rayleigh fading networks is presented in \cite{Prallel_relay} and
\cite{shahab}. The first capacity result for relay networks is
obtained in \cite{CoverRelay}, where the authors prove the
optimality of $\mathbf{DF}$ strategy in a single-relay network when
the received signal at the destination is a degraded version of the
relay received signal. Clearly, the degradedness condition holds in
the two-hop setting. Thus, $\mathbf{DF}$ would be the optimal
relaying scheme for two-hop networks. Indeed, most of the results in
the literature on relay networks either assume static channels
between the nodes or perfect knowledge of the Channel State
Information (CSI) at both end nodes of each link, for the case of
fading channels.

Recently, some papers discuss different transmission schemes over
relay networks when the CSI is not available at the transmitting
nodes, where most of them focus on Diversity-Multiplexing Trade-off
(DMT) \cite{laneman,azarian,yang_belfiore2,vkumar,shahabdmt,salman}.
Obviously, for such settings, the ergodic capacity is not defined,
however, the outage capacity is defined as the maximum rate
decodable with a given probability \cite{Outage_Def}. From the
throughput maximization point of view, the goal is to propose a
scheme to maximize the average data rate received at the
destination. The simplest form of such a problem is to find the
optimal transmission strategy for a one-hop single-user network when
CSI is available only at the receiver and the channel has
quasi-static fading characteristic. In a pioneering work, Shamai has
addressed this problem by substituting the receiver by a continuum
of virtual receivers, each corresponding to a specific realization
of the channel gain \cite{ShamaiBs}. Relying on the resulting
degraded broadcast channel, \cite{ShamaiBs} shows that an
infinite-level coding scheme with a proper power allocation among
the different levels of code maximizes the destination's average
data rate.

There are several extension for this work; in
\cite{Imper_csi_1,Imper_csi_2,Imper_csi_3} the authors try to find
the optimal transmission strategy when partial information is
available at the source. \cite{Mirghad} suggests the application of
multilevel coding in a multicast network with some QoS constraints
and derives the optimum throughput-coverage trade-off for such a
network. \cite{Multi_HARQ} combines multilevel coding scheme with
Hybrid Automatic Retransmission Request (HARQ) and shows that this
approach results in high throughput and low latency in a
point-to-point link. The Multiple Input Multiple Output (MIMO)
extension of \cite{ShamaiBs} is also discussed in
\cite{ShamaiBsMIMO}, \cite{Multi_steiner}, and
\cite{Vahid_Multi_MIMO}.

Considering the two-hop network with no CSI at the transmitter side
of each link, Steiner {\em et al} in \cite{ShamaiBsRelay} propose
different transmission schemes, assuming that the relay node has a
power constraint and is simple, i.e., is not capable of data
buffering over multiple coding blocks or rescheduling. To find the
optimal transmission scheme, they study different multilevel coding
schemes when the relay operates in different modes, including
$\mathbf{DF}$, $\mathbf{AF}$, and $\mathbf{CF}$. As discussed in
\cite{ShamaiBsRelay}, due to the high complexity of the
infinite-level $\mathbf{DF}$ coding scheme, they have only
considered a finite level code in their proposed $\mathbf{DF}$
strategies. Comparing the results of \cite{ShamaiBsRelay}, it turn
out that $\mathbf{AF}$ strategy outperforms all other investigated
transmission schemes, specifically in the high SNR regime. However,
as concluded in \cite{ShamaiBsRelay}, although $\mathbf{AF}$ has the
best performance among the other strategies, the optimality of
$\mathbf{AF}$ scheme is not implied. In fact, since the general
infinite-level $\mathbf{DF}$ coding scheme remains unsolved, there
is always a question of whether or not $\mathbf{DF}$ relaying
achieves a higher expected rate as compared to $\mathbf{AF}$. The
main motivation of this work is to answer this question.

In this work, we study the performance of $\mathbf{DF}$ relaying
scheme which uses infinite-level codes at the source and the relay
in the same setup as in \cite{ShamaiBsRelay}. To this end, we first
prove that the infinite-level code and $\mathbf{DF}$ relaying is
indeed the optimal two-hop transmission strategy for maximizing the
destination's average data rate. We further propose an algorithm to
determine the optimum power allocation for each indefinite-level
code. Numerical results are presented to verify that the proposed
scheme outperforms the $\mathbf{AF}$ strategy discussed
in~\cite{ShamaiBsRelay}.

The organization of this paper is as follows: First, section
\ref{sec:System model} describes the two-hop system model. A brief
review of some previous results on single-hop links are presented in
section \ref{sec:TransScheme}. Formulating the $\mathbf{DF}$
multilevel coding scheme in subsection \ref{ssec:DF_scheme}, we
prove the optimality of this scheme in subsection
\ref{sec:MultiOptimality}. Afterwards, in section
\ref{sec:TwoHopPower}, we present a procedure to actually determine
the optimal infinite-level code parameters at the source and the
relay. Numerical results and comparison with other schemes for the
special case where both links are Rayleigh fading is presented in
section \ref{sec:PerformAnal}. Finally, section \ref{sec:conclude}
concludes the paper.

Throughout this paper, we represent the expectation operation by
$E[.]$. The notation $\log (.)$ is used for the natural logarithm and
rates are expressed in nats. We denote $f_y(.)$ and $F_y(.)$ as the
probability density function (\textit{pdf}) and the cumulative distribution
function (CDF) of random variable $y$.

\section{System Model}\label{sec:System model}

\begin{figure}
\centering
\includegraphics[scale=0.45]{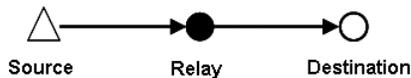}
\caption [Two-hop Network Model] {\small Two-hop Network Model
\normalsize} \label{fig:NetModel}
\end{figure}

In this paper, we investigate the performance of a two-hop network,
in which a  relay assists the transmission of data between a source
and a destination. As Fig.~\ref{fig:NetModel} shows, in a two-hop
network, the destination can solely receive data via the relay. All
nodes are assumed to have single antennas.
It is assumed that the source has no information about either of the
channel gains, the relay knows only the channel gain between itself
and the source, and the destination knows both channel gains. This
CSI assumptions are indeed practical, since the receiver of each hop
can evaluate its immediate channel gain by measuring the pilot
signal sent from the corresponding transmitter. In addition, the
receiver can measure the equivalent channel (the source to the
destination) if the relay forwards the pilot signal of the source
towards the destination. Having the equivalent channel gain and the
relay to destination channel gain, the destination can find the
source to relay channel gain as well.

It is also assumed that the source and relay both know the fading
power distribution of both links. By $f_\Gamma(\gamma)$ and
$f_\Upsilon(\upsilon)$ we denote the probability density function of
fading power in the first and second hop, respectively. The channels
of both links are assumed to be quasi-static, i.e., they are chosen
randomly (based on their corresponding \textit{pdf}) at the start of
the transmission and remain fixed for the whole transmission.

Source and relay are assumed to have power constraints $P_s$ and
$P_r$, respectively. Furthermore, they are assumed to be capable of
constructing infinite-level codes. However, the relay is assumed to
be simple, i.e., is not capable of buffering any unsent data,
rescheduling the untransmitted data from the previous blocks. It is
also assumed that the relay can not perform water-filling of its
power over multiple blocks and has a maximum power constraint in
each block. Therefore, it can only retransmit the data which it has
just received from the first hop. Successive decoding is used as the
decoding procedure at both destination and relay (in case the relay
wants to decode the data).

\section{Multilevel Coding Scheme for Single-hop Networks}\label{sec:TransScheme}

In this section, we review some previous studies on the application
of the multilevel codes for a single-hop network in two scenarios:
1) No CSI is available at the transmitter, which is known as
``broadcasting approach'' \cite{ShamaiBs} and 2) no CSI is available
at the transmitter and in addition, the transmission data rate can
not exceed a certain value \cite{ShamaiBsRelay}.

\subsection{Single-Hop Broadcasting Strategy} \label{ssec:OneHopTrans}

The optimum scheme for a single-hop link, in case that the
transmitter knows the fading power for each block is to design a
single-level code with the rate of $\log(1+lP)$ for that block.
Here, $P$ denotes the normalized transmission power, i.e, the
equivalent transmission power if the noise power is equal to one and
$l$ is the fading power for that specific block. Therefore, the
average achievable rate for this setup would be
$R_{erg}=E_l\left[\log(1+lP)\right]$ \normalsize
\cite{TelatarTechReport}.

Although designing a single-level code is optimum in the above
scenario, it can not be applied for a transmitter which does not
have access to the channel state information. For such a scenario,
Shamai has introduced a technique called broadcasting strategy
\cite{ShamaiBs}. In this technique, the transmitter sends the data
through infinite levels of a superposition code. Then, conditioned
on the channel state, i.e., the fading power, the receiver decodes
up to a certain level of the code. Therefore, the total receiving
rate for each channel realization, say $l$, can be evaluated as:
\begin{eqnarray}
\label{eq:R(l)}
        R(l)=\int\limits_{0}^{l}dR(a),
    \end{eqnarray}
\normalsize where $dR(a)$ represents the differential rate
transmitted over level `$a$' of the code. Defining $\rho(l)dl$ as
the power assigned for the $l^{th}$ level, $dR(l)$ is given by:
%\footnotesize
    \begin{eqnarray} \label{eq:dR(l)}
        dR(l) &=& \log \left(1+ \frac{l\rho(l)dl}{1+lI(l)}\right) \notag\\
&\simeq& \frac{l\rho(l)dl}{1+lI(l)},
    \end{eqnarray}
\normalsize where $I(l)=\int_{l}^{\infty}\rho(a)da$ and the second
line follows from the assumption of infinitesimal rate assignment
which is shown to be optimal in \cite{Mirghad} . The aim is to find
a $\rho(.)$ such that the average data rate at the destination is
maximized while the source power constraint ($P$) holds. It means
that the power should be assigned to different code levels such
that:
% \footnotesize
    \begin{eqnarray} \label{eq:OnehopObjFun}
        \max_{\rho(.)} && R_{av}=\int\limits_{0}^{\infty}dlf(l)R(l)\\
        \nonumber   s.t. && \int\limits_{0}^{\infty} \rho(l)dl =P,
    \end{eqnarray}
\normalsize where $f(.)$ shows the \textit{pdf} of the channel fading power.

This maximization problem has been studied and solved using calculus
of variations technique (see \cite{ShamaiBs} for details of the
proof). Here, we only mention the final solution as: %\footnotesize
    \begin{eqnarray} \label{eq:OnehopOpt}
        I^*(l)= \left\{ \begin{array}{l l}
            P & l<l_0 \\
            \frac{1-F(l)-lf(l)}{l^2f(l)} & l_0<l<l_1 \\
            0 & l_1<l
        \end{array}
        \right.,
    \end{eqnarray}
\normalsize where $F(l)=\int_{-\infty}^{l}f(a)da$ is the CDF of the
fading power. $l_0$ and $l_1$ are determined such that they satisfy
$I^*(l_0)=P$ and $I^*(l_1)=0$, respectively. Clearly, the optimum
power assignment can be determined by
$\rho^*(l)=-\frac{dI^*(l)}{dl}$. The maximum destination's average
data rate will be:
%\footnotesize
    \begin{eqnarray} \label{eq:OnehopAvg}
        R^*_{av}=\int_{l=0}^{\infty}dlf(l)R^*(l),
\end{eqnarray}
where $R^*(l)$ is obtained by setting $\rho (l) = \rho^* (l)$ in
(\ref{eq:R(l)}) and (\ref{eq:dR(l)}). The total rate
of the final superimposed code can be evaluated by: %\footnotesize
    \begin{eqnarray} \label{eq:OnehopFeed}
        R^*_{F}=\int_{0}^{\infty}dR^*(l).
    \end{eqnarray}
Note that this value only depends on the \emph{fading power
distribution} and \emph{the transmitter power}.

Finally, it is important to mention that the above scheme is indeed
the \textit{optimal transmission scheme} for a fading link when the
source does not have the CSI. It is due to the result of
\cite{Bergmans} which proves that the multilevel coding maximizes
any weighted sum-rate of a degraded broadcast channel. Therefore,
considering the equivalent broadcast model for the single-user
fading channel, the multilevel coding achieves the optimal
destination's  average data rate.
\subsection{Single-Hop Rate-Limited Broadcast Strategy}
\label{ssec:OneHopTransLimitRate}

An interesting extension of the above broadcast strategy is
designing a multilevel code  for a source with available data rate
limited to $R_{in}$, i.e., the transmission  rate should be less
than $R_{in}$ \cite{ShamaiBsRelay}.
%This limitation can be implied by the user traffic
%model or packet dropping due to some network congestions.

The problem formulation is similar to subsection
\ref{ssec:OneHopTrans}, except it has one more constraint on the
transmission rate. The modified
optimization problem will be, \cite{ShamaiBsRelay}: %\footnotesize
    \begin{eqnarray} \label{eq:RateLimObjFun}
        \max_{\rho(.)} && R_{av}=\int\limits_{0}^{\infty}dlf(l)R(l)\\
        \nonumber s.t. && \int\limits_{0}^{\infty} \rho(l)dl =P, ~~ \mbox{and}~~\int\limits_{0}^{\infty} dR(l) \leq R_{in},
    \end{eqnarray}
\normalsize where the second condition ensures that the transmission
rate remains less than the available source data rate.

Reference \cite{ShamaiBsRelay} uses \emph{constrained} calculus of
variations to solve this problem. More precisely, initially the
authors in \cite{ShamaiBsRelay} substitute the first constraint by
two end-point conditions of $I(0)=P$ and $I(\infty)=0$. Then, using
$\rho^*(l)=-\frac{dI^*(l)}{dl}$ and Lagrangian multiplier,
\eqref{eq:RateLimObjFun} is reconstituted as a variations problem
(See \cite{ShamaiBsRelay} for more details). It turns out that the
optimum distribution function is as follows: %\footnotesize
    \begin{eqnarray} \label{eq:RateLimOpt}
        I^*(l)= \left\{ \begin{array}{l l}
            P & l<l_0 \\
            \frac{1-F(l)+\lambda-lf(l)}{f(l)l^2} & l_0<l<l_1 \\
            0 & l_1<l
        \end{array},
        \right.
    \end{eqnarray}
where $F(l)=\int\limits_{-\infty}^{l}f(a)da$. $l_0$ and
$l_1$ are determined as a function of $\lambda$ and
satisfy $I^*(l_0)=P$ and $I^*(l_1)=0$, respectively. Finally, $\lambda$ is
computed such that the transmission rate constraint holds.

As an example, for the case of Rayleigh fading channel, i.e.,
$F(l)=1-e^{-l}$, the optimum distribution is as follows
\cite{ShamaiBsRelay}: %\footnotesize
    \begin{eqnarray} \label{eq:RateLimOpt1}
        I^*(l) &= & \left\{ \begin{array}{l l}
            P & l<l_0 \\
            \frac{\lambda}{e^{-x}x^2}+\frac{1}{x^2}-\frac{1}{x}
            & l_0<l<l_1 \\
            0 & l_1<l
        \end{array}
        \right. \\
        \nonumber \\
        l_1&=&1-W_L(-\lambda{e})
    \end{eqnarray}
\normalsize where $W_L(x)$ is the Lambert W-function and
finds $w$ such that $we^w=x$. The values for $\lambda$ and $l_0$ are
also determined by solving the following system of equations:
%\footnotesize
    \begin{eqnarray} \label{eq:RateLimOpt2}
    \left\{\begin{array}{l l l}
        R_{in}&=&2\log(l_1)-l_1-(2\log(l_0)-l_0)\\
        I^*(l_0)&=&P
        \end{array}
        \right.
    \end{eqnarray}
\normalsize

One important observation is  that in the case of $R_{in} \geq
R^*_F$ ($R^*_F$ is defined in equation (\ref{eq:OnehopFeed})) the
above rate-limited problem will be simplified to the original
problem in subsection \ref{ssec:OneHopTrans}. Although this
statement can be verified mathematically, its intuitive explanation
would be insightful. It is obvious that if $R_{in}$ tends to
infinity, the rate condition is always satisfied (will not be an
active constraint). Therefore, maximization problem of
\eqref{eq:RateLimObjFun} relaxes to the form of equation
\eqref{eq:OnehopObjFun}. Moreover, \ref{ssec:OneHopTrans} shows that
the source, in the optimal transmission of the original problem
(without rate constraint),  feeds the channel with a rate equal to
$R^*_F$. Hence, it turns out that even though the available rate at
the source is more than $R^*_F$, the source needs to transmit only
$R^*_F$ bits of information in each block. Therefore, if $R_{in}
\geq R^*_F$, the solutions of the two optimization problems of
\eqref{eq:OnehopObjFun} and \eqref{eq:RateLimObjFun} are equal.

\begin{figure}
\centering
\includegraphics[scale=0.40]{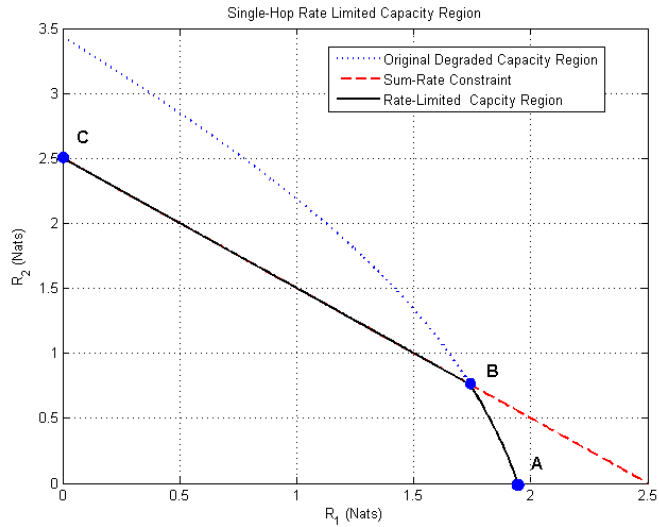}
\caption [Single-Hop Rate Limited Capacity Region] {\small
Single-Hop Rate Limited Capacity Region  \normalsize}
\label{fig:BC_Cap}
\end{figure}

At the end, note that the  resulted multilevel code is also the
\emph{ optimal transmission scheme} for the rate-limited one-hop
set-up. To prove, we should use the broadcast equivalent structure
of the network. Indeed, the capacity region of the rate-limited
degraded broadcast network is the intersection of the original
degraded broadcast channel capacity region and the region  below the
surface associated with the transmission rate constraint. For
illustration, the solid line in Fig. \ref{fig:BC_Cap} shows a
typical capacity region of a rate-limited two-user degraded
broadcast channel. In this figure, dotted and dashed lines depict
the capacity region of the original degraded broadcast channel
(without rate limitation) and the line representing the rate
constraint, respectively. The destination's average data rate can be
evaluated as the sum of the received data rate for each channel
state $l$ ($R(l)$) times the probability of occurring that specific
channel state ($f(l)$). Clearly, this value is maximized on the
boundary of the resulted capacity region (the solid line).
Therefore, the optimal point would be either over the capacity
region without rate limitation (Arc $\overset{\frown}{AB}$) or the
end point of the rate limitation line (Point \textbf{C}). Since
$\overset{\frown}{AB}$ is a part of the original capacity region,
multilevel coding is the optimal scheme to achieve any point on
$\overset{\frown}{AB}$. Furthermore, point \textbf{C} is archived
using a single level code which is again a special case of
multilevel codes. This proves the optimality of multilevel coding
scheme for rate limited scenarios.

\section{Multilevel Coding Scheme for Two-hop Networks}\label{sec:TwoHopTrans}

As an extension of the one-hop set-up, \cite{ShamaiBsRelay}
addresses the problem of maximizing the destination's average data
rate in a two-hop network, where there is no direct link between the
source and the destination. In reference \cite{ShamaiBsRelay},
several schemes have been studied, including broadcasting strategy
with $\mathbf{AF}$ relaying, and $\mathbf{DF}$ relaying with finite
level broadcasting at the source and the relay. Infinite level codes
with $\mathbf{DF}$ relaying is also addressed in
\cite{ShamaiBsRelay}. However, the performance of this method
remains as an open problem there.

In this section, we will first describe the infinite level
$\mathbf{DF}$ strategy in details. Then, in subsection
\ref{sec:MultiOptimality}, we prove the optimality of this scheme.
Finally, subsection \ref{sec:TwoHopPower} presents an algorithm to
optimally design such an infinite level code.

\subsection{$\mathbf{DF}$ Infinite-Level Codes for Two-hop
Networks} \label{ssec:DF_scheme} Based on the system model, each
transmission block of the infinite level $\mathbf{DF}$ strategy
consists of the following two steps:

\begin{enumerate}
  \item In the first phase,
  the source allocates its power among
  different code levels with the power distribution function $\rho_s(.)$.
  Of course, $\rho_s(.)$ should satisfy the power constraint $\int_{0}^{\infty} \rho_s(a)da =P_s$.
  Then, based on the source-relay channel fading power,
  say $\gamma$, the relay is able to decode up to the level $\gamma$
  of the transmitted data. Thus, the relay received rate is:
%  \footnotesize
  \begin{eqnarray} \label{eq:TwohopRelayRate}
      R_r(\gamma)=\int\limits_{0}^{\gamma} \log \left(1+ \frac{a\rho_s(a)da}
      {1+aI_s(a)} \right) \simeq
      \int\limits_{0}^{\gamma} \frac{a\rho_s(a)}{1+aI_s(a)}da,
  \end{eqnarray}
  \normalsize where $I_s(a)=\int_{a}^{\infty}\rho_s(a)da$.

  \item In the second phase, the relay should transmit the
  data to the destination. As noted earlier, in this work, we only focus
  on \emph{simple relays}
  which \emph{can  neither buffer any of the previously received data nor do any scheduling
  tasks}. As a results, these relays have two features which
  seem obvious but have important effects on the code design.
  To illustrate, consider a case in which the relay has decoded
  $R_r(\gamma)$ bits of the transmitted data.
  It turns out that,
  firstly, the relay can not transmit with the rate greater than
  $R_r(\gamma)$.
  Secondly, if the relay transmits with the rate $R_2$, $R_2<R_r(\gamma)$,
  the rest of the data ($R_r(\gamma)-R_2$) can not be stored and should
  be discarded.
  Consequently, the relay, in each transmission block, should
  choose the optimal power distribution of the multilevel code such
  that it satisfies the relay total
  power constraint ($P_r$). Meanwhile, the relay should keep the transmission
  rate below its received data rate in that block ($R_r(\gamma)$).

  Defining
  $\rho_r(.|R_r(\gamma))$ as the
  power distribution of  each code level at the relay
  conditioned on the input rate of $R_r(\gamma)$, we can
  summarize these conditions as:
  \begin{enumerate}
    \item Power constraint at the relay:
    $\forall R_r(\gamma):$\\
    $\int\limits_{0}^{\infty}\rho_r(a|R_r(\gamma))da=P_r$.
    \item Available rate constraint at the relay:
    $\forall R_r(\gamma):$\\
    $\int\limits_{0}^{\infty}\frac{a\rho_r(a|R_r(\gamma))da}
    {1+aI_r(a|R_r(\gamma))}
    \leq R_r(\gamma)$,   where  $R_r(\gamma)$  is defined by \eqref{eq:TwohopRelayRate}.
  \end{enumerate}
  Clearly, the relay requires to know $\rho_r(a|R_r(\gamma))$ for all
  possible values of $R_r(\gamma)$.
  %in order to optimally assign its power depending on different realizations of the
  %first hop.

  Transmitting a multilevel code on the relay-destination
  link, the destination is able to
  decode up to a certain level `$\upsilon$'. Here, `$\upsilon$' denotes the
  fading power of the second link. Therefore, for each $R_r(\gamma)$,
  the received data rate at the  destination can be written as: %\footnotesize
  \begin{eqnarray} \label{eq:TwohopDestRate}
    R_d(\upsilon|R_r(\gamma))=\int\limits_{0}^{\upsilon} \log \left(1+\frac{a\rho_r(a|R_r(\gamma))da}{1+aI_r(a|R_r(\gamma))} \right)
        \simeq \int\limits_{0}^{\upsilon}\frac{a\rho_r(a|R_r(\gamma))}{1+aI_r(a|R_r(\gamma))}da .
  \end{eqnarray} \normalsize
  Indeed, for successful decoding of the signal, the destination should know the
  power allocation strategy of the relay. This information can be
  obtained through the knowledge of the source to the relay channel gain.

\end{enumerate}

\vspace{.3cm}
 Given these, we are now able to formulate the two-hop
optimization problem. Similar to the single-hop scenario, we want to
maximize the average data rate received at the destination. Assuming
$f_\Gamma(\gamma)$ and $f_\Upsilon(\upsilon)$ as the probability
density functions of the fading power in the source-relay and
relay-destination links, respectively, the
destination's average data rate can be written as: %\footnotesize
\begin{eqnarray} \label{eq:TwohopERd}
    E[R_d]&=&E_\Gamma\left\{E_\Upsilon[R_d\left(\upsilon|R_r(\gamma)\right)]\right\}\\
      \nonumber &=&\int\limits_{0}^{\infty}
    \int\limits_{0}^{\infty}
     f_\Gamma(\gamma)f_\Upsilon(\upsilon)\int\limits_{0}^{\upsilon}\frac{a\rho_r(a|R_r(\gamma))}
     {1+aI_r(a|R_r(\gamma))}dad\upsilon~d\gamma.
\end{eqnarray} \normalsize
Therefore, we obtain the final optimization
problem as follows:  %\footnotesize
\begin{eqnarray} \label{eq:TwohopOptProblem1}
    \max_{\rho_s(.),~~\rho_r(.)} && \int\limits_{0}^{\infty}
    \int\limits_{0}^{\infty}
    f_\Gamma(\gamma)f_\Upsilon(\upsilon)R_d(\upsilon|R_r(\gamma))d\upsilon~d\gamma\\
    \nonumber s.t. &&  \int\limits_{0}^{\infty} \rho_s(a)da =P_s, \\
    \nonumber    &&
    \forall R_r(\gamma): \int\limits_{0}^{\infty}\rho_r(a|R_r(\gamma))da=P_r,\\
    \nonumber    &&
    \forall R_r(\gamma): \int\limits_{0}^{\infty}
    \frac{a\rho_r(a|R_r(\gamma))da}{1+aI_r(a|R_r(\gamma))} \leq R_r(\gamma).
\end{eqnarray}
\normalsize Note that the above optimization problem is similar to
the one derived in \cite{ShamaiBsRelay}. However, in
\cite{ShamaiBsRelay}, the last constraint (relay rate limitation) is
stated as an equality. In fact, it is more accurate to formulate the
rate limitation by an inequality constraint instead of equality. It
is due to the fact that the relay may not have to send all
information it receives from the first hop to achieve  the optimal
performance. In other words, the last constraint of equation
\eqref{eq:TwohopOptProblem1} lets the relay to discard some of its
received data if it wants to do so. For instance, this may happen
when the relay receives data rate higher than its corresponding
$R^*_F$ ($R^*_F$ is defined in equation (\ref{eq:OnehopFeed})). In
such a scenario, the relay only uses $R^*_F$ bits of the received
information and ignores the rest.

\subsection{Optimality of Two-Hop $\mathbf{DF}$ Multilevel Coding}
\label{sec:MultiOptimality}

The main focus of this section is to show that the multilevel coding
approach combined with decode and forward ($\mathbf{DF}$) relaying
 maximizes the average data rate at the destination of a two-hop
network.

To start, let us emphasize that according to the two-hop structure
of the network, all information received by the destination  should
be first passed through the relay and there is not any direct link
between the source and the destination. As a result, the destination
received signal is always a degraded version of what has been
received at the relay. In other words, no information can be decoded
by the destination unless it has been decodable at the relay. Thus,
it can be concluded that decode and forward is the optimal relaying
scheme for two-hop settings.

\begin{figure}
\centering
\includegraphics[scale=0.60]{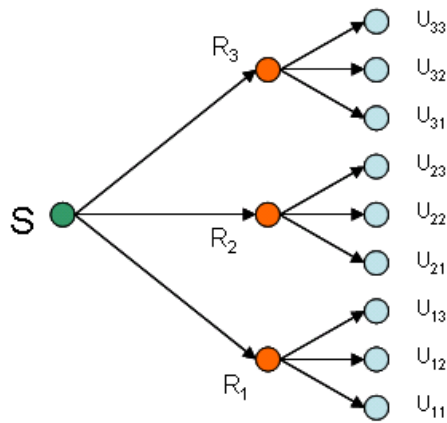}
\caption [Two-Hop Equivalent Broadcast Channel Model] {\small
Virtual Two-Hop Broadcast Channel Model, $\kappa=3$ \normalsize}
\label{fig:VirtualTwo}
\end{figure}

Knowing the optimality of $\mathbf{DF}$, similar to the single-hop
network (See \cite{ShamaiBsMIMO}), we model both first and second
fading hops by infinite number of virtual relays and virtual users,
respectively. More precisely, we substitute the relay with infinite
relays, each has a constant channel gain which corresponds to a
specific realization of the first hop channel. These virtual relays
constitute a degraded set. If we assume that the channel fading
power is selected from a set of discrete values
$\{a_1,a_2,\cdots,a_\kappa\}$\footnote{Note that here, for
simplicity, we assume that the fading levels of both links are
selected from the same set. However, the statements of optimality
holds for the general case. Moreover, this assumption is valid for
the case of $\kappa \to \infty$ }, there would be $\kappa$ virtual
relays in the network. Without loss of generality, we assume
${a_1<a_2<\cdots<a_\kappa}$. Of course, this model is accurate only
if $\kappa$ tends to infinity. In a similar way, the second hop can
be modeled with $\kappa$ virtual users for each of the virtual
relays; thus, in total there would be $\kappa^2$ virtual users in
the network. $U_{ij}$ denotes the virtual user which is associated
to the virtual relay $R_i$, and its channel gain is $a_j$. Moreover,
$B_{ij}$ represents the decodable rate at this node. For
illustration purpose, Fig. \ref{fig:VirtualTwo} depicts the network
model for the case that $\kappa=3$, i.e., $A=\{a_1,a_2,a_3\}$.

The aim is to find the  \emph{optimum transmission scheme}  for the
source and the relay. We start from the second hop and assume the
source uses an arbitrary transmission scheme. Furthermore, let
$\{B_1,B_2,\cdots,B_\kappa\}$ be the rate that each of the virtual
relays, i.e., $\{R_1,R_2,\cdots,R_\kappa\}$, can decode under this
transmission scheme. Having the data rate $B_{i}$ at relay $R_i$,
the optimal second-hop strategy is to maximize the destination's
average data rate for each of the virtual relays. As will be
described in subsection \ref{sec:TwoHopPower}, the solution to this
problem is similar to the result of the subsection
\ref{ssec:OneHopTransLimitRate} and the optimal power distribution,
$\rho^*_{B_i}(u)$, for each of the virtual relays can be determined
by \eqref{eq:RateLimOpt}, in which $R_{in}=B_i$. In other words,
multilevel coding is the optimal strategy for the second hop.

Note that the successful decoding is only possible if the
destination knows the applied $\rho^*_{B_i}(u)$ (or equivalently the
value of $B_i$) at the relay. In our two hop model, this information
can be obtained by the knowledge of the source to the relay channel
gain at the destination. The optimal received data rate for each of
the virtual users, say $U_{il}$, can be evaluated by:
\begin{eqnarray} \label{eq:Optim_Sec_hop}
    R^*(B_i,\Upsilon=a_l)=\int\limits_{0}^{a_l}\frac{u\rho^*_{B_i}(u) du}{1+uI^*_{B_i}(u)}
\end{eqnarray}
where $I^*_{B_i}(u)=\int_{u}^{\infty}\rho^*_{B_i}(a)da$. Let
$\psi_{\Upsilon}(j)$ be probability that the second hop gain is
equal to $a_j$. Hence, the optimal average rate of the second hop
under rate condition $B_i$ would be:
\begin{eqnarray} \label{eq:Optim_G}
    D^*(B_i)=\sum\limits_{l=1}^{\kappa} \psi_{\Upsilon}(l)R^*(B_i,\Upsilon=a_l).
\end{eqnarray}
Given $D^*(B_i)$,  the two-hop network average data rate can be
written as:
\begin{eqnarray} \label{eq:Optim_R}
    \overline{R}=\sum\limits_{l=1}^{\kappa} \psi_{\Gamma}(l)D^*(B_l),
\end{eqnarray}
where $\psi_{\Gamma}(l)$ represents the probability that the first
hop channel is in  state $a_l$. The goal of the code design is to
maximize $\overline{R}$. As \eqref{eq:Optim_R} shows, the
destination's average data rate is the weighted sum of a non-linear
function of $B_i$. The domain of acceptable $B_i$ for
$i\in\{1,2,...,\kappa\}$ is a convex set which is known as the
capacity region of the underlying broadcast channel. Moreover, due
to the degradedness of the virtual relays, all points on the
boundary of this capacity region can be achieved by a multilevel
code \cite{CoverRelay}. Therefore, to prove the optimality of the
multilevel code for the first hop, it is enough to show that
$\overline{R}$ is maximized on the boundary of this capacity region.
This argument can also be justified if we can show that
$\frac{\partial \overline{R}}{\partial B_i}$ is positive $\forall
i\in\{1,2,...,\kappa\}$. To show this, we write:
\begin{eqnarray} \label{eq:Diff_R}
    \frac{\partial \overline{R}}{ \partial B_i}= \psi_{\Gamma} (i){D^*}'(B_i)
\end{eqnarray}
where $D^*(B_i)$ is defined in equation \eqref{eq:Optim_G} and shows
the destination's average data rate when the relay has the rate
constraint of $B_i$. By definition, $\psi_{\Gamma} (i)$ is a
non-negative value. Moreover, from the results of subsection
\ref{ssec:OneHopTransLimitRate}, it can be concluded that
${D^*}'(B_i)$ is non-negative. In fact, if $B_i\geq~R^*_{F}$, where
$R^*_{F}$ is defined in equation (\ref{eq:OnehopFeed}), then
${D^*}'(B_i)=0$ and ${D^*}'(B_i)>0$, otherwise. Therefore,
$\frac{\partial \overline{R}}{ \partial B_i}\geq~0$ for $\forall
i\in\{1,2,...,\kappa\}$. Consequently, $\overline{R}$ is maximized
for the values of $\{B_i\}_{i=1}^{\kappa}$ which are on the boundary
of the capacity region of the first hop broadcast channel. In other
words, the multilevel coding scheme is the optimal transmission
strategy for the first hop. This completes the proof for the
optimality of multilevel coding for the two-hop network.

\subsection{Optimal Design of Two-Hop $\mathbf{DF}$ Multilevel
Coding} \label{sec:TwoHopPower}

Having the optimality of two-hop multilevel coding scheme, in this
section, we present a procedure in order to solve the two-hop
optimization problem introduced in equation
\eqref{eq:TwohopOptProblem1}. The main difficulty of this problem is
that, unlike single-hop scenarios (the original and the rate-limited
broadcasting cases), equation \eqref{eq:TwohopOptProblem1} can not
be directly solved by variations methods. It is due to the fact that
the constraint on the second hop rate does not have a fixed value on
the right hand side, i.e., it does not have a form of isoperimetric
problem. For a complete discussion on isoperimetric problem, refer
to \cite{CalVar}. To solve this problem, we use the following lemma.
\begin{lem}\label{lem:Lemma1}
The following maximization problems on two non-negative functions $f_1(.)$ and $f_2(.)$:
\begin{eqnarray} \label{eq:Lemma}
    \max_{f_1(.),f_2(.)} \int_{x=a}^{b}\int_{y=c}^{d}dxdy\mathcal{H}(x)\mathcal{K}\left(f_1(.),f_2(.),x,y\right),
\end{eqnarray}
and
\begin{eqnarray} \label{eq:Lemma2}
    \max_{f_1(.)}
    \int_{x=a}^bdx\mathcal{H}(x)\max_{\{f_2(.)|(f_1(.),x)\}}\int_{y=c}^{d}\mathcal{K}\left(f_1(.),f_2(.),x,y\right)dy ,
\end{eqnarray}
where $\mathcal{H}(.)$ and $\mathcal{K}(.)$ are two known
non-negative functions, are equivalent.
\end{lem}
\begin{proof}
Let us denote the solution of (\ref{eq:Lemma}) by $\left(f_1^* (.),
f_2^* (.) \right)$ and the solution of (\ref{eq:Lemma2}) by
$\left(f_1^{\star} (.), f_2^{\star} (.) \right)$. We can write
\small
\begin{eqnarray} \label{ghar1}
\int_{x=a}^{b}\int_{y=c}^{d}dxdy\mathcal{H}(x)\mathcal{K}\left(f_1^*(.),f_2^*(.),x,y\right) &=& \int_{x=a}^{b} dx\mathcal{H}(x) \int_{y=c}^{d} \mathcal{K}\left(f_1^*(.),f_2^*(.),x,y\right) dy \notag\\
&\leq&  \int_a^bdx\mathcal{H}(x)\max_{\{f_2(.)|(f_1^*(.),x)\}}\int_{y=c}^{d}\mathcal{K}\left(f_1^*(.),f_2(.),x,y\right)dy \notag\\
&\leq& \max_{f_1(.)}
    \int_a^bdx\mathcal{H}(x)\max_{\{f_2(.)|(f_1(.),x)\}}\int_{y=c}^{d}\mathcal{K}\left(f_1(.),f_2(.),x,y\right)dy.
\end{eqnarray}
\normalsize On the other hand, \small
\begin{eqnarray} \label{ghar2}
\max_{f_1(.)}
    \int_a^bdx\mathcal{H}(x)\max_{\{f_2(.)|(f_1(.),x)\}}\int_{y=c}^{d}\mathcal{K}\left(f_1(.),f_2(.),x,y\right)dy &=& \int_{x=a}^{b} dx\mathcal{H}(x) \int_{y=c}^{d} \mathcal{K}\left(f_1^{\star}(.),f_2^{\star}(.),x,y\right) dy \notag\\
&=& \int_{x=a}^{b}\int_{y=c}^{d}dxdy\mathcal{H}(x)\mathcal{K}\left(f_1^{\star}(.),f_2^{\star}(.),x,y\right) \notag\\
&\leq& \max_{f_1(.),f_2(.)} \int_{x=a}^{b}\int_{y=c}^{d}dxdy\mathcal{H}(x)\mathcal{K}\left(f_1(.),f_2(.),x,y\right). \notag\\
\end{eqnarray}
\normalsize
Combining (\ref{ghar1}) and (\ref{ghar2}), lemma is
proved.
\end{proof}

Using this lemma and noting that $f_\Gamma(\gamma) \geq 0, \forall
\gamma$ , we can reform $E[R_d]$ in (\ref{eq:TwohopOptProblem1}) as
follows:
\begin{eqnarray} \label{eq:TwohopOptProblemReArreng}
     \max_{\rho_s(.),\rho_r(.)} E[R_d] &=&
      \max_{\rho_s(.)}  \int\limits_{0}^{\infty} d\gamma~f_\Gamma(\gamma)
      \max_{\left\{\rho_r(.)|\rho_s(.),\gamma\right\}} \int\limits_{0}^{\infty}d\upsilon~f_\Upsilon(\upsilon)
      \int\limits_{0}^{\upsilon}\frac{a\rho_r(a|R_r(\gamma))}{1+aI_r(a|R_r(\gamma))}da,\nonumber\\
      &\hspace{-0.3cm}\stackrel{(a)}{=}\hspace{-0.3cm}&
      \max_{\rho_s(.)}  \int\limits_{0}^{\infty} d\gamma~f_\Gamma(\gamma)
      \max_{\rho_r(.|R_r(\gamma))}\int\limits_{0}^{\infty}d\upsilon~f_\Upsilon(\upsilon)
      \int\limits_{0}^{\upsilon}\frac{a\rho_r(a|R_r(\gamma))}{1+aI_r(a|R_r(\gamma))}da,
\end{eqnarray}
\normalsize where the outer maximization is subject to: %\footnotesize
\begin{eqnarray} \label{eq:TwohopOptProblemOuterCons}
    \int\limits_{0}^{\infty} \rho_s(a)da =P_s,
\end{eqnarray}
\normalsize and the constraints of the inner problem are as follows:
%\footnotesize
\begin{eqnarray}
   & \forall R_r(\gamma): \int\limits_{0}^{\infty}
   \rho_r(a|R_r(\gamma))da=P_r, &\\
   & \forall R_r(\gamma): \int\limits_{0}^{\infty}
   \frac{a\rho_r(a|R_r(\gamma))}{1+aI_r(a|R_r(\gamma))}da \leq  R_r(\gamma). &
   \label{eq:TwohopOptProblemInerCons}
\end{eqnarray}
In \eqref{eq:TwohopOptProblemReArreng}, $(a)$ follows from the fact
that $R_r(\gamma)$ can be determined with the knowledge of $\gamma$
and $\rho_s(.)$ and the dependence of the term $\int\limits_{0}^{\infty}d\upsilon~f_\Upsilon(\upsilon)
      \int\limits_{0}^{\upsilon}\frac{a\rho_r(a|R_r(\gamma))}{1+aI_r(a|R_r(\gamma))}da$ on $\rho_s (.), \gamma$ is only through $R_r(\gamma)$.

Given
\eqref{eq:TwohopOptProblemReArreng}-\eqref{eq:TwohopOptProblemInerCons},
in the following two subsections, we will discuss how this two-step
maximization problem can be solved using Euler's equations
\cite{CalculeVar}.

\subsubsection{Relay-Destination Link Optimization Problem}
\label{ssec: InerMax}

Receiving $R_r(\gamma)$ bits from the first hop, the aim of the
relay  is to maximize the average data rate received at the
destination. In fact, if the input rate changes, the relay should
modify its power distribution, accordingly. However, the knowledge
of the input rate ($R_r(\gamma)$), the relay total power, and the
\emph{pdf} of the second hop fading power is sufficient for
determining the optimum power distribution function,
$\rho^*_r(.|R_r(\gamma))$. It is evident that the optimum power
distribution function, $\rho^*_r(.)$, can be completely determined
by evaluating $\rho^*_r(.|R_r(\gamma))$ for all values of
$R_r(\gamma)$. $\rho^*_r(.|R_r(\gamma))$, itself, is the
solution of the following problem: %\footnotesize
\begin{eqnarray} \label{eq:TwohopInnerOptProb}
    h(R_r(\gamma)) & \triangleq& \max_{\rho_r(.|R_r(\gamma))} \int\limits_{0}^{\infty}d\upsilon~f_\Upsilon(\upsilon)
    \int\limits_{0}^{\upsilon}\frac{a\rho_r(a|R_r(\gamma))da}{1+aI_r(a|R_r(\gamma))} \\
    && s.t.  \quad \int\limits_{0}^{\infty} \rho_r(a|R_r(\gamma))da=P_r,\notag\\
    && \quad \quad  \int\limits_{0}^{\infty}
   \frac{a\rho_r(a|R_r(\gamma))}{1+aI_r(a|R_r(\gamma))}da \leq  R_r(\gamma).\notag
\end{eqnarray}
\normalsize Note that, in \eqref{eq:TwohopInnerOptProb},
$R_r(\gamma)$ is a constant; hence, the problem takes the form of
the rate-limited broadcast strategy problem, subsection
\ref{ssec:OneHopTransLimitRate}. Therefore, the optimum solution is:
%\footnotesize
    \begin{eqnarray} \label{eq:InnerOptSol}
        I^*_r(l|R_r(\gamma))= \left\{ \begin{array}{l l}
            P_r & l<l_0 \\
            \frac{1-F_\Upsilon(l)+\lambda-lf_\Upsilon(l)}{f_\Upsilon(l)l^2} & l_0<l<l_1 \\
            0 & l_1<l
        \end{array}
        \right. ,
    \end{eqnarray}
\normalsize where
$F_\Upsilon(l)=\int\limits_{-\infty}^{l}f_\Upsilon(a)da$. $l_0$ and
$l_1$ are determined as a function of $\lambda$ to
 satisfy $I^*_r(l_0)=P_r$ and $I^*_r(l_1)=0$, respectively. The
optimum multilevel power distribution at the relay can be found by
$\rho_{r}^*(l| R_r(\gamma))=-\frac{dI^*_{r}(l|R_r(\gamma))}{dl}$.
Finally,
$\lambda$ is computed to satisfy: %\footnotesize
\begin{eqnarray} \label{eq:TwohopInnerOptProbRate}
    \int\limits_{0}^{\infty}
    \frac{a\rho_r(a|R_r(\gamma))}{1+aI_r(a|R_r(\gamma))}da= \min(R_r(\gamma), R^*_F),
%    \frac{a\rho_r(a|R_r(\gamma))}{1+aI_r(a|R_r(\gamma))} \leq R_r(\gamma))
\end{eqnarray}
\normalsize where $R^*_F$ is defined by \eqref{eq:OnehopFeed}. This
condition comes from the fact that achieving the maximum average
rate at the destination requires the  relay not to transmit more
than $R^*_F$ bits (refer to the discussion in subsection
\ref{ssec:OneHopTransLimitRate}).

\begin{figure}
\centering
\includegraphics[scale=0.40]{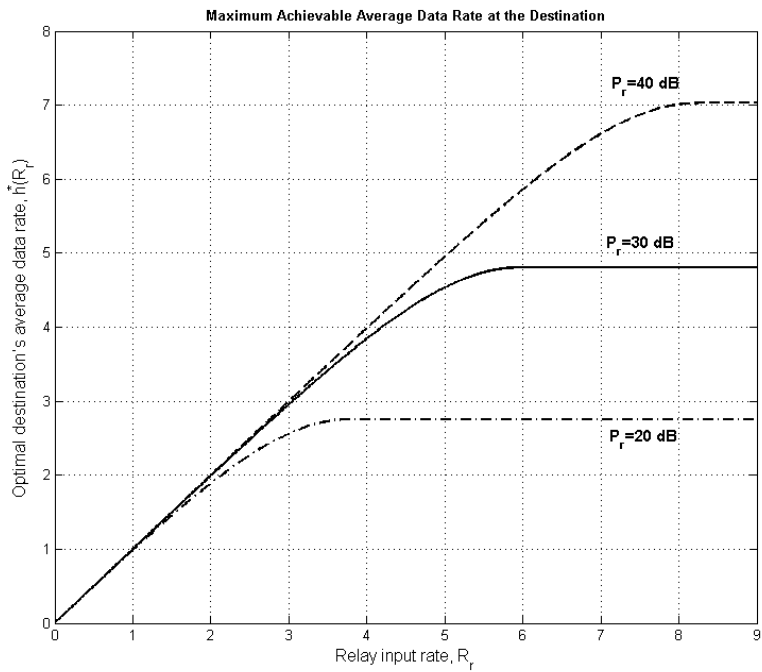}
\caption [Maximum Average Data Rate Received at the Destination ]
{\small Maximum Average Data Rate Received at the
Destination}%, $P_r$=20, 30 \& 40 dB }
\normalsize \label{fig:hstarval}
\end{figure}

As an example, we have solved \eqref{eq:TwohopInnerOptProb} for a
network in which the second hop can be modeled as a Rayleigh fading
channel, i.e., $F_\Upsilon(\upsilon)=1-e^{-\upsilon}$.
Fig.~\ref{fig:hstarval} shows the maximum destination's  average
data rate, $h^*(R_r(\gamma))$, for different relay input rates
($R_r(\gamma)$) and different relay powers ($P_r$).

\subsubsection{Source-Relay Link Optimization Problem} \label{ssec:
OuterMax}

Knowing the optimum value for the inner integration,
$h^*(R_r(\gamma))$, \eqref{eq:TwohopOptProblemReArreng} can be
written as follows:
%\footnotesize
\begin{eqnarray} \label{eq:OuterOpt}
    E[R_d]=\max_{\rho_s(.)} && \int\limits_{0}^{\infty} d\gamma~f_\Gamma(\gamma)
    h^*(R_r(\gamma))\\
    \nonumber s.t. &&  \int\limits_{0}^{\infty} \rho_s(a)da =P_s,
\end{eqnarray}
\normalsize where: %\footnotesize
\begin{eqnarray} \label{eq:Rrdef}
        R_r(\gamma)= \left\{ \begin{array}{l l}
            0 & \gamma<l_0 \\
            \int_{l_0}^{\gamma} \frac{a\rho_s(a)}{1+aI_s(a)}da & l_0<\gamma<l_1 \\
            \int_{l_0}^{l_1} \frac{a\rho_s(a)}{1+aI_s(a)}da & l_1<\gamma
        \end{array}
        \right.,
\end{eqnarray}
\normalsize and $I_s(l)=\int\limits_{l}^{\infty}\rho_s(a)da$. $l_0$
and $l_1$ satisfy $I_s(l_0)=P_s$ and $I_s(l_1)=0$, respectively. As
\eqref{eq:Rrdef} suggests, $R_r(\gamma)$ only depends on $\gamma$,
$\rho_s(.)$, and $I_s(.)$. Remembering
$\rho_s(l)=-\frac{dI_{s}(l)}{dl}$, we can write the integrand of
\eqref{eq:OuterOpt} as
$G(l,I_s,I'_s)=f_\Gamma(l)h^*(R_r(l,I_s,I'_s))$. With this notation,
equation \eqref{eq:OuterOpt} takes the form of a fixed end-point
Calculus of Variations problem and can be solved using
Euler's equation, \cite{CalculeVar},%\footnotesize
\begin{eqnarray} \label{eq:EulerLag}
    \zeta(l,I_s,I'_s)=G_{I_s}-\frac{dG_{I'_s}}{dl}=0,
\end{eqnarray}
\normalsize where $G_{I_s}=\frac{\partial G}{\partial
R_r}\frac{\partial R_r}{\partial I_s}$,  $G_{I'_s}=\frac{\partial
G}{\partial R_r}\frac{\partial R_r}{\partial I'_s}$, and
$\frac{dG_{I'_s}}{dl}$ is the derivative of $G_{I'_s}$ with respect
to $l$. Thus, we have:

%\footnotesize
\begin{eqnarray} \label{eq:EulerLagConGIs}
    G_{I_s}&=&\frac{\partial G}{\partial R_r}\frac{\partial R_r}{\partial I_s}
       =  \left\{ \begin{array}{l l}
            f_\Gamma(l)h^{*'}(R_r(l,I_s,I'_s)) \int\limits_{0}^{l} \frac{a^2I'_s(a)}{(1+aI_s(a))^2}da & l_0<l<l_1 \\
            0 & \mbox{\emph{otherwise}} \\
        \end{array}
        \right.,
\end{eqnarray}
\begin{eqnarray} \label{eq:EulerLagConGIPS}
    G_{I'_s}&=&\frac{\partial G}{\partial R_r}\frac{\partial R_r}{\partial I'_s}
           =  \left\{ \begin{array}{l l}
            f_\Gamma(l)h^{*'}(R_r(l,I_s,I'_s)) \int\limits_{0}^{l} \frac{-a}{1+aI_s(a)}da & l_0<l<l_1 \\
            0 & \mbox{\emph{otherwise}} \\
            \end{array}
            \right.,\\
            \nonumber \\
    \frac{dG_{I'_s}}{dl}&=&\left\{ \begin{array}{l l} \label{eq:EulerLagConGIPSoverL}
                \begin{array}{l}
                f'_\Gamma(l)h^{*'}(R_r(l,I_s,I'_s)) \int\limits_{0}^{l} \frac{-a}{1+aI_s(a)}da+f_\Gamma(l)h^{*'}(R_r(l,I_s,I'_s))\frac{-l}{1+lI_s(l)}\\
                ~~~~~~~~~~~~~~~~~~~+f_\Gamma(l)h^{*''}(R_r(l,I_s,I'_s))\frac{-lI'_s(l)}{1+lI_s(l)} \int\limits_{0}^{l} \frac{-a}{1+aI_s(a)}da
                 \end{array} & l_0<l<l_1 \\
                 \\
            0 & \mbox{\emph{otherwise}} \\
            \end{array}
            \right.,
\end{eqnarray}
\normalsize
where $h^{*'}(.)$ and $h^{*''}(.)$ denote the first order and the second order derivative of $h^*(.)$, respectively.
Substituting
\eqref{eq:EulerLagConGIs}-\eqref{eq:EulerLagConGIPSoverL} in
\eqref{eq:EulerLag}, the optimal $I_s^*(l)$ is derived.

\begin{figure}
\centering
\includegraphics[scale=0.45]{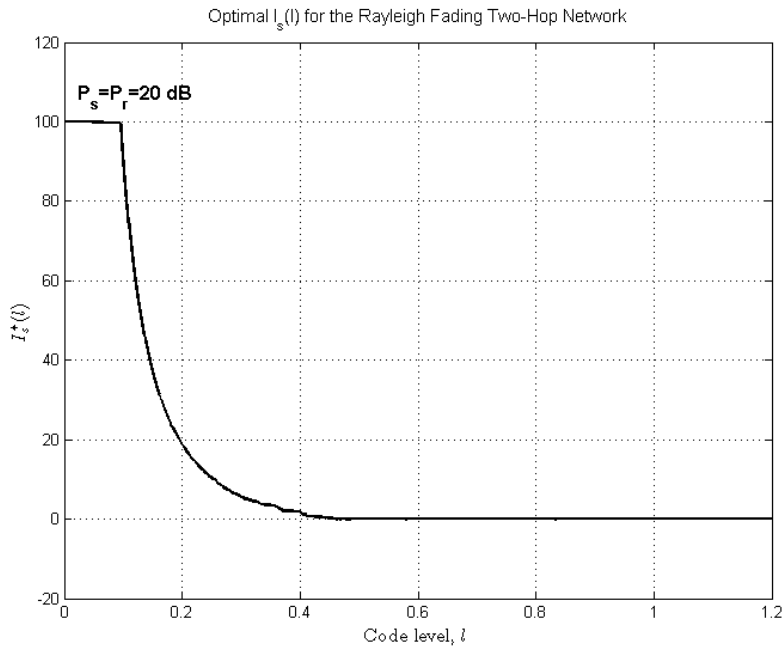}
\caption [Optimal $I_s(l)$ for the Rayleigh Fading Two-Hop Network ]
{\small Optimal $I_s(l)$ for the Rayleigh Fading Two-Hop Network
\normalsize} \label{fig:Isopt}
\end{figure}

As an example, in  the scenario where both source-relay and
relay-destination links are modeled with a Rayleigh fading channel,
i.e., $F_\Gamma(l)=F_\Upsilon(l)=1-e^{-l}$, \eqref{eq:EulerLag}
can be simplified to: %\footnotesize
\begin{eqnarray} \label{eq:RayleighZeta}
    \zeta(l,I_s,I'_s)&=&h^{*'}(i)\left[\int\limits_{0}^{l} \frac{1-a-a^2I_s(a)}{(1+aI_s(a))^2}da\right]\\
      \nonumber  && -h^{*''}(i)\left[\frac{-lI'_s(l)}{1+lI_s(l)}\int\limits_{0}^{l}
      \frac{-a}{1+aI_s(a)}da\right]=0,
\end{eqnarray}
\normalsize where $i=R_r(l,I_s,I'_s)$. To solve
(\ref{eq:RayleighZeta}), we first need to have $h^{*'}(i)$ and
$h^{*''}(i)$. Indeed, these functions can be numerically evaluated
using the results of subsection \ref{ssec: InerMax}. In the next
step, we replace $I_s$ by $\left[I_s(1),I_s(2),...,I_s(N)\right]$,
corresponding to the amount of interference in each
level\footnote{In fact, we have approximated a continuous variable
$I_s(l)$ with a discrete $N$-level function, which becomes precise
as $N$ tends to infinity.}. $I_s (m)$'s are in descending order,
such that $I_s(1)=P_s$ and $I_s(N)=0$. As a result, we have a
nonlinear system of $N$ equations, i.e., $\zeta(m,I_s,I'_s)=0, \,
\,m=\{1,2,...,N\}$ which can be solved numerically. The final
solution for these $N$ variables shows the optimal interference
function, $I^*_s(l)$.
%need Revise
As an example, Fig. \ref{fig:Isopt} presents $I^*_s(l)$ in the case
of
 Rayleigh fading model for both hops and $P_s=P_r=20$dB. Having $I^*_s(l)$,
the amount of power associated for  each code level can be
determined by $\rho^*_s(l)=\frac{-dI^{*}_{s}(l)}{dl}$.
\normalsize

\newpage
\section{Numerical Results and Comparison with Other Schemes} \label{sec:PerformAnal}

In the previous sections, we have proposed a $\mathbf{DF}$
multilevel coding scheme, which is shown to be optimum in the
underlying network setup, and derived the optimum source and relay
power distribution through different levels of code. In this
section, we compare the performance of the proposed  scheme (which
is the optimal scheme) with the cut-set upper-bound and two other
sub-optimal schemes proposed in \cite{ShamaiBsRelay}.
% and briefly explained in the following

\renewcommand{\labelenumi}{\Alph{enumi}.}
\begin{enumerate}
  \item \textbf{Broadcasting Cutset Bound}, $\mathbf{C_{cutset}}$ :\\ This
  bound simply says that the  achievable average data rate of a two-hop network
  can not exceed the achievable average rate of any of the single-hop links,
  i.e., the source-relay and the relay-destination links. This is independent of
  the relay structure and its operation. In other words, the cutset bound is an upper-bound
  on the network throughput when we put no limitation on the relay, i.e., the relay is capable
  of buffering the unsent data or rescheduling the buffered data. Therefore, the gap between the
    performance of the proposed scheme and the cutset upper-bound
    shows the maximum possible gain of having a ``complicated''
    relay instead of a simple one. The cutset bound
  can be written as:%\footnotesize
    \begin{eqnarray} \label{eq:cutbound}
        C_{cutset}=\min\left[\int_0^\infty~da~f_\Gamma(a)R_1(a),\int_0^\infty~da~f_\Upsilon(a)R_2(a)\right],
    \end{eqnarray}
    \normalsize
    where $R_1(a)$ ($R_2(a)$) denotes the rate that the relay (destination) can successfully
    decode when the source (relay) transmits over a channel with fading power equal to `$a$'.

  \item \textbf{Amplify and Forward}, $\mathbf{AF}$:\\
    This is the achievable rate of a two-hop network
    in which the relay performs the amplify and forward ($\mathbf{AF}$) on the
    received source signal. To design the optimum
    multilevel power distribution, first, the
    total equivalent channel should be evaluated. In other words, the
    source-relay and relay-destination channels combined with $\mathbf{AF}$ relaying
    can be modeled as one channel with a new
    probability density function. Having this new \emph{pdf}, the
    optimum power distribution can be evaluated. Details of the proof
    can be found in \cite{ShamaiBsRelay}. The final result can be written
    as:
    %\footnotesize
    \begin{eqnarray} \label{eq:AF}
        R_{AF}=\int_{l_0}^{l_1}da\left[ \frac{2(1-F_{s_b}(a))}{a}+\frac{(1-F_{s_b}(a))f'_{s_b}(a)}{f_{s_b}(a)}\right],
    \end{eqnarray}
    \normalsize where:   %\footnotesize
    \begin{eqnarray} \label{eq:AF2}
        F_{s_b}(x)=1-\int_{\frac{P_s}{P_r}a}^{\infty}dl~e^{-l-\frac{a(1+P_rl)}{lP_r-aP_s}},
   \end{eqnarray}
   \normalsize
   and $f_{s_b}(a)=\frac{d}{da}F_{s_b}(a)$. $l_0$ and $l_1$ are defined such that
   $I_{opt}(l_0)=P_s$ and $I_{opt}(l_1)=0$, respectively. $I_{opt}(l)$
   is presented in \cite{ShamaiBsRelay}.

\newpage
  \item \textbf{Outage at the Source, Broadcasting at the Relay}, $\mathbf{DF_{1-bs}}$:\\
        This scheme is another suboptimal strategy that has
        been studied in \cite{ShamaiBsRelay}. In this case, the source uses a one level
        code, known as the \emph{outage approach}, and the relay uses the optimal
        multilevel code. The subscript ``$\mathbf{1-bs}$'' represents the one-level
        coding and the broadcast scheme at the source and relay,
        respectively.
        Clearly, this approach is a special case of the optimum $\mathbf{DF}$ broadcast strategy, i.e., the
        proposed scheme. The achievable average rate of this scheme
        can be computed by:
        %\footnotesize
        \begin{eqnarray} \label{eq:DF1bs1}
        R_{DF,1-bs} &=& \max_{\gamma_s,\rho_r(.)}  (1-F_\Gamma(\gamma_s))\int\limits_{0}^{\infty}da(1-F_\Upsilon(a))\frac{a\rho_r(a)}{1+aI_r(a)}\\
        \nonumber s.t. &&  \int\limits_{0}^{\infty} \frac{u\rho_r(u)du}{1+uI_r(u)}=\log(1+P_s\gamma_s).
        \end{eqnarray}
        \normalsize
\end{enumerate}

\begin{figure}
\centering
\includegraphics[scale=0.45]{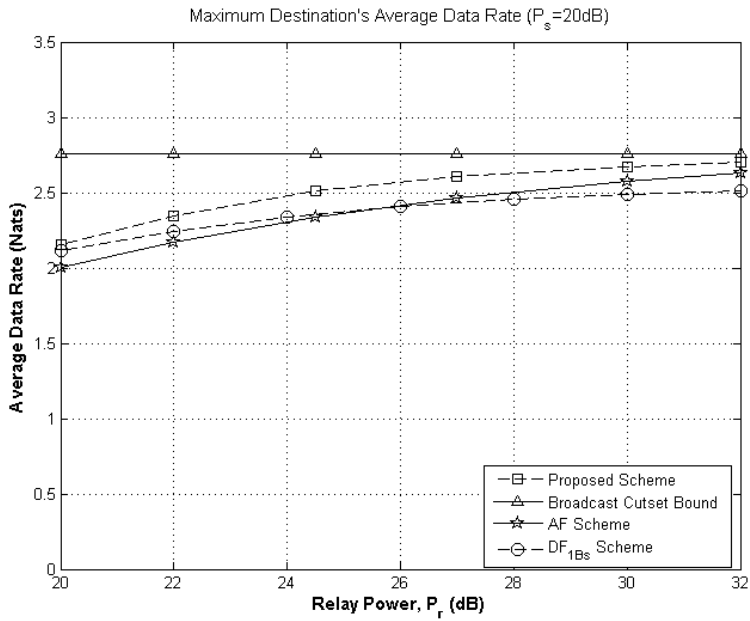}
\caption [Destination's Average Data Rate, $P_s$=20dB] {\small
Destination's Average Data Rate $P_s$=20~dB, $P_r$=20 - 32~dB
\normalsize} \label{fig:Final20}
\end{figure}

\begin{figure}
\centering
\includegraphics[scale=0.45]{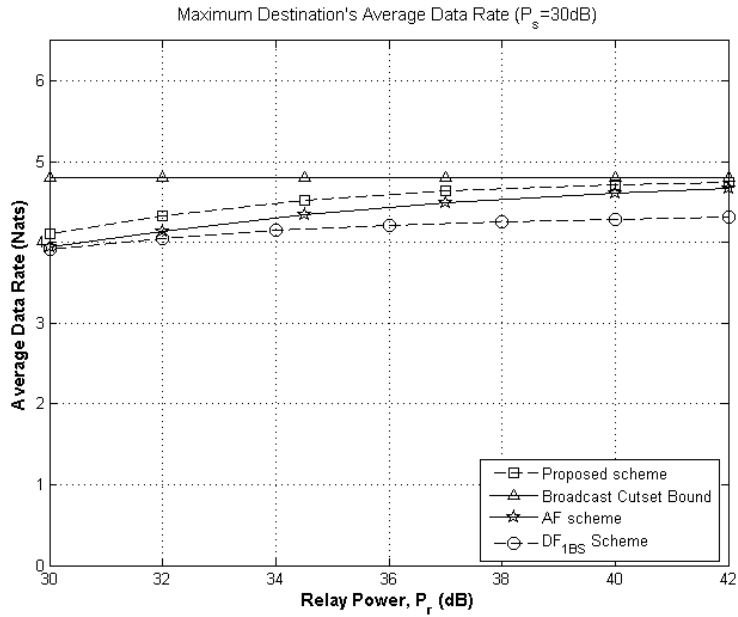}
\caption [Destination's Average Data Rate, $P_s$=30dB] {\small
Destination's Average Data Rate $P_s$=30~dB, $P_r$=30 - 42~dB
\normalsize} \label{fig:Final30}
\end{figure}

Figures (\ref{fig:Final20}) and (\ref{fig:Final30}) represent the
destination's average data rate at the destination versus the relay
power $P_r$ for the proposed scheme, as well as the $\mathbf{AF}$
and $\mathbf{DF_{1-bs}}$  schemes, where $P_s=20$dB and $P_s=30$dB,
respectively. The upper-bound $\mathbf{C_{cutset}}$ is also depicted
in both figures.

As expected, the proposed $\mathbf{DF}$ strategy (the optimal
scheme) outperforms the $\mathbf{AF}$ and $\mathbf{DF_{1-bs}}$
schemes. Note that, the superiority of the proposed scheme over
$\mathbf{DF_{1-bs}}$ is obvious since  $\mathbf{DF_{1-bs}}$ is a
special case of the proposed scheme. The important observation in
these figures is that the infinite multilevel $\mathbf{DF}$ strategy
is strictly superior to the $\mathbf{AF}$ strategy, which was
previously the best known scheme for this setup at the high SNR
\cite{ShamaiBsRelay}. However, as the SNR at the relay side $P_r$
increases, the performance of $\mathbf{AF}$ approaches the optimal
performance. Furthermore, as $P_r$ increases, the proposed scheme
approaches the cutset bound which means that for high values of
$P_r$ the relay does not need to be ``complicated''. Another
observation from these figures is that, as $P_r$ decreases,
$\mathbf{DF_{1-bs}}$ approaches the optimal performance. This can be
explained as follows: when the power of the relay is much smaller
than the source power, the relay-destination link limits the
performance. Therefore, even using a one-level code at the source is
sufficient to deliver an average rate of $R^*_F$ to the relay, which
is the maximum rate that relay can transmit to the
destination\footnote{Note that even if relay receives a rate more
than $R^*_F$, the extra rate should be discarded.}.

\vspace{-.2cm}
\section{Conclusion} \label{sec:conclude}
In this paper, a two-hop network in which the data is transmitted
from the source node via a single relay to a destination node was
considered. It was assumed that the knowledge of the channel for
each hop is not available at the corresponding transmitter. The
relay was assumed to be simple, i.e., not capable of data buffering
over multiple coding blocks, water-filling over time, or
rescheduling. For this network setup, we proposed an infinite-level
coding scheme at the source and the relay. It is shown that this
scheme in conjunction with the Decode and Forward ($\mathbf{DF}$)
relaying is indeed the optimal strategy for maximizing the average
data rate received at the destination. We also proposed an algorithm
to find the optimum amount of power which should be assigned to each
code level at the source and relay. The optimality of the
$\mathbf{DF}$ multilevel coding strategy is also verified through
numerical results by showing its superiority over the Amplify and
Forward ($\mathbf{AF}$) scheme, which was previously the best known
scheme for the high SNR regime.

% Index
% Put a \makeindex command in the Preamble if you use MakeIndex program
% and put
%???????????????????????????????????????
%\printindex % here
%???????????????????????????????????????
% OR, do it "by hand" inside \begin{theindex} ... \end{theindex}
%----------------------------------------------------------------------
\renewcommand{\baselinestretch}{1.1}
\bibliographystyle{IEEE} % sorted alphabetically, labeled with numbers
%\bibliography{keylatex} % names file keylatex.bib as my bibliography file

%\bibliographystyle{plain}
\bibliography{keylatex}

\end{document}